\documentstyle[aps,epsf]{revtex}  
\newcommand{\mnu}{\mathrm m_{\nu_{\tau}}}

\newcommand{\tautoc}{\tau \to 5 \pi^{\pm} \nu_{\tau}}
\newcommand{\tautos}{\tau \to 5 \pi^{\pm} \pi^{0} \nu_{\tau}}
\newcommand{\tautot}{\tau \to 3 \pi^{\pm} \nu_{\tau}}
\newcommand{\ehmh}{M_{h},E_{h}}
\begin{document}        
\baselineskip 14pt
\title{Determination of the upper limit on $\mnu$ from LEP.}
\author{Fabio Cerutti}
\address{Laboratori Nazionali dell'INFN Frascati, Via E. Fermi 40, 00044 Frascati Italy\\
CERN EP Division, 1211 Geneva 23, Switzerland }
%
\maketitle              

\begin{abstract}        
A review of the direct determinations of the upper limit on the
tau-neutrino mass from the LEP experiments is given. The experimental methods,
the results and the comparison with non LEP measurements are also 
discussed. The study of the systematic errors shows that the LEP 
results are statistically limited so that their combination will improve 
the sensitivity to a massive tau-neutrino. An unofficial combination of 
the ALEPH and OPAL measurements gives a 95 $\%$ confidence level 
upper limit of 15 MeV/c on $\mnu$.
\end{abstract}          

\section{Introduction: Motivation and indirect constraints.}

The neutrino masses are one of the most puzzling and hot subject
of discussion in the high energy physics community.
It is believed that the smallness the neutrinos masses
can be explained by assuming that they are produced by the
mixing between {\it standard} Dirac mass terms and large Majorana 
mass terms; the Majorana masses are related to a new energy scale at which 
the lepton number conservation is 
violated. This is the so called {\it see-saw}~\cite{seesaw} 
mechanism which is present in many grand-unified models. 
If the Standard Model mass hierarchy is preserved in the 
Dirac sector of the neutrino mass matrix the  
tau-neutrino is expected to be by far the heaviest  
neutrino. Under this assumption the neutrino mass hierarchy is 
expected to be of the order 
$m_{\nu \tau}:m_{\nu \mu}:m_{\nu e} = m_{t}^2:m_{c}^2:m_{u}^2$. 

Cosmology~\cite{cosmology} can put strong 
constraints on the neutrino masses because of their influence on 
the actual density of the universe. An unstable 
tau-neutrino with a mass of the order of 10-20 MeV can survive 
to the cosmological constraints.
Measurements of light nucleus abundances which results 
from Big Bang Nuleo Synthesis~\cite{BBNS} can give information 
on the neutrino masses. The incompatibility between the BBNS prediction 
and the measured D and H$^{3}$ abundances can be solved by and unstable tau-neutrino 
with a mass of the order of 10-25 MeV~\cite{kawasaki}.

The {\it claimed} superK~\cite{superk} discovery of atmospheric neutrinos 
oscillation would constrain $\mnu$ to be lighter than about 170 KeV (which is 
the direct limit on the mu-neutrino mass) if what they 
observe is an oscillation between tau and mu neutrinos.

\section{The fit to $\mnu$: 2-dimensional method}
The $\mnu$ measurements at LEP are based on a fit to 
the $E_{h},M_{h}$ spectrum in hadronic tau decays. This method has been 
introduced for the first time by two LEP experiments OPAL~\cite{opal5}
and ALEPH~\cite{aleph5}. The fact that for each given hadronic 
mass the hadronic energy is constrained between the two values 
$E_{h}^{max,min}(M_{h},\mnu)$ gives a sizable improvement in 
the sensitivity to the tau-neutrino mass with respect to the one 
obtained by a fit to the $M_{h}$ spectrum alone, as explained 
in ~\cite{fabio,luca}.

The two decays used at LEP are $\tautoc$ and $\tautot$. The first decay mode 
benefits of a invariant mass spectrum which extends to large values 
but is limited by the very small branching which is of the order 
of 0.08$\%$. 
The second mode benefit of a large statistics 
(BR($\tautot$)$\sim 9 \%$) but the $M_h$ spectrum is suppressed at 
values close to $M_{\tau}$ because of the $a_1$ dominance in 3$\pi$ tau decays. 
The two modes have similar sensitivities 
even though the regions in the $\ehmh$ plane from which this sensitivity comes are 
slightly different: for the five-prong mode it comes from few events at 
very high values of $M_{H}$ and $E_{h}$ while for the three-prong mode events at 
very high energy and intermediate mass can contribute too. 

The value of $\mnu$ is obtained by a likelihood fit to
the observed events where the likelihood has the following form:

\begin{eqnarray}
{\cal L}(m_{\nu}) = \prod_{events}^{ }
 \frac {1} {\Gamma} \cdot 
 \frac {d^{2} \Gamma} {dE_{h} dm_{h}}  
 \otimes {\cal G} (E_{beam},E_{\tau})
 \otimes {\cal R} (m_{h},E_{h},\rho,\sigma_{m_{h}},\sigma_{E_{h}},...) 
 \otimes \varepsilon(m_{h},E_{h})
\end{eqnarray}

The  $\frac {d^{2} \Gamma} {dE_{h} dm_{h}}$
is the double-differential tau decay width and contains the unknown 
part related to the hadronic spectral functions. The knowledge of these 
function is relevant in estimating the sensitivity of an experiment
(the so called {\it luck factor}) but it doesn't affect the limit 
which comes from a region of the $\ehmh$ plane where the phase space is 
dominant (this is true if no narrow resonance is present with 
$M_{res} \sim M_{\tau}$ as discussed in the three-prong 
results section). The effect of initial/final state
radiation is described by ${\cal G}(E_{beam},E_{\tau})$; at 
LEP ISR is expected to be small and it has practically very 
small effect on the $\mnu$ determination. 

The resolution function  
${\cal R} (m_{h},E_{h},\rho,\sigma_{m_{h}},\sigma_{E_{h}},...)$ is 
the most delicate part of this measurement; the determination of 
the $\ehmh$ end-point requires the knowledge of
the tracking calibration with an accuracy 
better than the ratio $\mnu / M_{tau}$. As will be shown in the result 
section this is the main source of systematics for all the LEP 
experiments. The detector efficiency is contained in the function 
$ \varepsilon(m_{h},E_{h})$; since this function is not expected to 
vary rapidly in the sensitive region its influence on the $\mnu$ 
determination is expected to be very small.

\section{The Results}

In this section the LEP results from five- and three-prong tau 
decays are reviewed. In the three-prong section the possible problem 
caused by the presence of a narrow resonance close to the hadronic
mass end-point is discussed. Finally the ALEPH and OPAL results are 
combined with the likelihood product method by using the published
five- and three-prong likelihoods. An estimate of the systematic 
error of the combined result is given too. 

\subsection{Results form $\tautoc$ tau decays}

The decay $\tautoc$ decay mode has been used by ALEPH~\cite{aleph5}
and by OPAL~\cite{opal5} to measure the tau-neutrino mass. Both 
experiments have analysed the full LEP1 statistics which 
corresponds to about 200k  tau-pairs. 

\begin{figure}[h]      
\centerline{\epsfxsize 2.0 truein \epsfbox{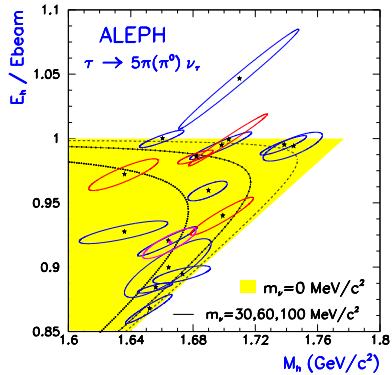}}   
\vskip 0.2 cm
\caption[]{
\label{aleph5}
\small The five-prong ALEPH events in the $\ehmh$ mass planes. 
The iso-lines corresponding to the kinematic region allowed for 
a 30, 60 and 100 MeV tau-neutrino are superimposed.}  
\end{figure}

The ALEPH experiment
has selected 52 $\tautoc$ decays (and 3 $\tautos$ decays which 
due to the worse $\pi^{0}$ energy resolution have very small
impact on the final result) with an efficiency of about 27$\%$ 
and a background from {\it dangerous} topologies at the level of 0.6$\%$. In 
terms of the $\mnu$ upper limit the {\it dangerous} backgrounds are the tau 
decays in which the hadronic mass and/or the hadronic energy are reconstructed 
at values larger than the true ones. For example a decay 
$\tau \to 3\pi^\pm \pi^0 \to  3\pi^\pm \gamma e^+ e^-$ 
where the two electrons are reconstructed as pions tends to have 
a reconstructed hadronic mass which is systematically higher than the 
true one. If this events are not rejected they could mimic a massless
tau-neutrino giving a {\it fake}  good limit on $\mnu$. The
same problem holds for $q \bar q$ events reconstructed as $\tautoc$; 
in fact this kind of events tends to be in the high $\ehmh$ region. 
The typical ALEPH 
resolutions are of about 15 MeV for $M_{h}$ and 350 MeV for $E_{h}$.
The resolution parameters have been determining by using the so called 
Monte Carlo {\it cloning} technique~\cite{aleph5} which allows the 
determination of these parameters on an event by event basis. 

The fit to the ALEPH events showed in Fig.\ref{aleph5} gives a limit of 
$\mnu < 22.3$ MeV at 95$\%$ confidence level. The systematic error 
is dominated by the knowledge of the parameters of the resolution 
function. The energy and the mass scales and resolutions have been 
determined by using the $Z \to \mu^{+} \mu^{-}$ events and the 
charm decays $D^{0} \to K^{-} \pi^{+}$, $D^{0} \to K^{-} \pi^{+} \pi^{+} \pi^{-}$
and $D^{+} \to K^{-} \pi^{+} \pi^{+}$. By adding linearly the 
0.8 MeV systematic error the final 95 $\%$ C.L. limit is 
$\mnu < 23.1$ MeV. 

\begin{figure}[hb]      
\centerline{\epsfxsize 2.0 truein \epsfbox{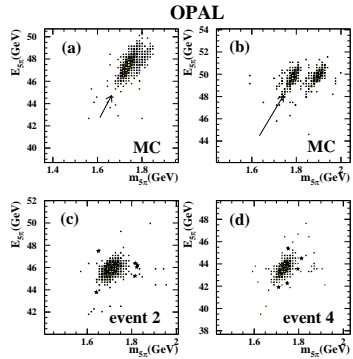}}   
\vskip 1.0 cm
\caption[]{
\label{opalclone}
\small The results of the Monte Carlo cloning technique applied 
to two Monte Carlo events with reconstruction problems are shown 
in figures (a) and (b); the beginning of the arrow shows the generated 
values of $M_h$ and $E_h$, the end of the arrow shows values of these 
quantities after the first Monte Carlo reconstruction and the boxes 
show the distribution of these quantities obtained by applying the 
Monte Carlo {\it cloning} technique to these two events. 
The result of the same technique 
applied to the two most sensitive data events is shown in figures (c) 
and (d).}
\end{figure}

The OPAL experiment has performed a similar 
measurement by selecting 22 $\tautoc$ decays~\cite{opal5}. The selection 
efficiency is of 9.3$\%$ with a {\it dangerous} background of the order of 
2.5$\%$. The parameters of the resolution function have been determined, as for 
ALEPH,  
with the Monte Carlo {\it cloning} technique. In the OPAL paper is 
proved that this technique is able to spot events with reconstruction 
problem as shown in Fig.~\ref{opalclone}.

Typical mass and energy resolutions of the OPAL analysis are 20-25 MeV and 500 
MeV respectively. The fit to the 22 OPAL events gives a limit of $\mnu < 39.6$ 
MeV at 95$\%$ confidence level.
As for ALEPH the systematic error is dominated by the knowledge 
of the resolution function parameters and is of 3.6 MeV. By adding linearly this 
systematics to statistical limit OPAL obtains a 95 $\%$ C.L. upper limit on 
$\mnu$ of 43.2 MeV.

\subsection{Results form $\tautot$ tau decays}
As mentioned in the introduction the three-prong tau decay mode is competitive with
the five-prong one in the determination of the tau neutrino mass. The three LEP 
experiments ALEPH~\cite{aleph3}, DELPHI~\cite{delphi3} and OPAL~\cite{opal3} have
used this decay mode to constraint the tau-neutrino mass. 

The ALEPH results is based on a fit to the $\ehmh$ distribution. 
Due to the large statistics the fit has been limited 
to an high $\ehmh$ region where 3000 $\tautot$ decays have been selected. 
The selection efficiency in this region is of about 
49$\%$ with a background from {\it dangerous} topologies of less then 0.2$\%$.
The high statistics of this channel make the {\it cloning} technique not viable. 
For this reason ALEPH has parameterised the quantities entering in
resolution ${\cal R}$ as a function of the hadronic mass and energy. 
The typical values of the mass and of the energy resolution are similar to the 
ones obtained in the five-prong mode. 
The fit gives a statistical limit of 21.5 MeV on the tau-neutrino 
mass at 95$\%$ confidence level. The systematic error on this limit is again dominated by 
the knowledge of the resolution function and ammount to 4.2 MeV. This error is
larger than the five-prong one mainly because of the non use of the {\it cloning}
technique.
Adding linearly the systematic error to the fit result a 95$\%$ C.L.  limit of 
$\mnu < 25.7$ MeV has been obtained. 

The OPAL experiment has tried to increase its sensitivity to the tau neutrino mass
by partially reconstructing the tau direction in three-prong versus three-prong 
tau events.
In this kind of events the thrust axes is a good approximation of the tau 
direction especially for events where the three-prong are very energetic. By 
a fit to the two variables square missing-mass and missing-energy  on a sample of 
2514 events OPAL obtained an upper limit of 32.1 MeV at 95$\%$ C.L. on $\mnu$. 
The systematic error has been estimated to be of 3.2 MeV dominated by the 
knowledge of the resolution  function parameters. This gives a final limit
of $\mnu < 35.3$ MeV at 95 $\%$ confidence level.

The DELPHI experiment has selected 12538 $\tautot$ decays with a 38$\%$ efficiency
and a {\it dangerous} background of 1.5 $\%$. A fit to the $\ehmh$ distribution
gives a 95 $\%$ C.L. upper limit of 25 MeV on the tau-neutrino mass.
In the study of the systematics DELPHI has observed a significant 
disagreement between the three-prong mass spectrum in the data sample 
and the one obtained with a Monte Carlo based on the 
K\"un-Satamaria model~\cite{KS}.
This discrepancy is observed in the $M_h$ range (1.5-1.9) GeV. The DELPHI
collaboration claimed~\cite{delphiap} that this excess could be explained if about 
2.3$\%$ of new 
resonance, the a'(1700) with a mass of 1.7 GeV and a width of 0.3 GeV,
was added in the three-prong tau decay. The description to the Dalitz plots in the 
three-prong tau decays also improved by the addition of this resonance.
The CLEO experiment has tried to measure the 
ammount of this resonance in their three-prong tau sample~\cite{cleoap} 
(by assuming a massless tau-neutrino) and has obtained (with different models) an 
a'(1700) fraction of the order of (0.1-0.4)$\%$ which is significantly smaller than
the 2.3$\%$ reported by DELPHI.

The ALEPH and the OPAL experiments has observed the same problem as DELPHI 
in describing the $\tautot$ Dalitz plots however the do not observe any excess 
with respect to
the Kh\"un and Santamaria model in the hadronic mass spectrum.
The ALEPH experiment has checked the effect of such a large ammount of a'(1700) on its
limit: if a 2.5$\%$ of a'(1700) with the parameters suggested by DELPHI is added in 
the three-prong fit the limit on $\mnu$ is worsened by about 6 MeV. This implies 
a variation on the combined three- and five-prong ALEPH upper limit, reported in the 
following, of about 1 MeV.

As DELPHI correctly states that a simultaneous fit of the a'(1700) 
properties and of $\mnu$ in 
three-prong tau decays is not possible. In view of the CLEO results 
and of the ALEPH check is unlikely that the limit on the tau neutrino 
mass can be deteriorated by the presence of this new resonance. More inputs from 
theorists is welcome.

\subsection{Combination of the ALEPH and OPAL results}
The ALEPH and the OPAL collaborations have 
combined~\cite{aleph3,opal5} (separately) their three- and 
five-prong upper limits on the tau-neutrino masses. 
The method used to combine these 
results is based on the likelihood product. In doing the combination the correlation 
between the systematic errors of the two decay modes has been properly taken into 
account as described in~\cite{aleph3,opal5}. 
The limit obtained by the OPAL and by the ALEPH collaborations are respectively  of 
$\mnu < 27.6$ MeV  and of  $\mnu < 18.2$ MeV at 95$\%$ confidence level, including 
systematic effects. 

\begin{figure}[h]      
\centerline{\epsfxsize 2.0 truein \epsfbox{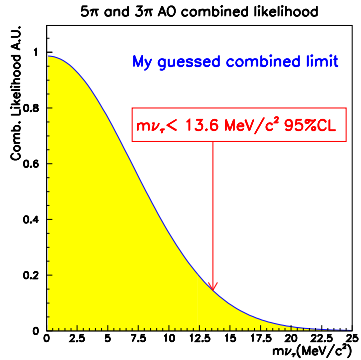}}   
\vskip 1.0 cm
\caption[]{
\label{likcomb}
\small The ALEPH and OPAL combined likelihood as a function of $\mnu$ 
obtained from five- and three-prong published results. This likelihood 
does not include systematic effects.}
\end{figure}

The results from the different LEP experiments and from the different tau decay modes 
are limited by statistics. Moreover the dominant systematics (resolution function
parameters) are mainly uncorrelated between the different LEP experiments.
For this reason a combination of the LEP results would improve the sensitivity to the 
tau-neutrino mass. I have done this exercise in order to get an estimate of what this 
combined limit would be. I have used the five- and three-prong likelihoods published 
by the ALEPH and by the OPAL experiments (the DELPHI results have not yet been 
published). The method is the same as the one used in the ALEPH and OPAL 
publications: 
\begin{eqnarray}
{\cal L}_{COMB}(m_{\nu}) =  {\cal L}_{OPAL}^{3\pi}(m_{\nu}) \times  {\cal L}_{OPAL}^{5\pi}(m_{\nu})
\times  {\cal L}_{ALEPH}^{5\pi}(m_{\nu}) \times  {\cal L}_{ALEPH}^{3\pi}(m_{\nu})
\end{eqnarray}
the combined likelihood is shown in Fig.~\ref{likcomb}.
 
From this likelihood a 95$\%$ C.L. limit $\mnu < 13.6$ MeV can be derived
by requiring $ln({\cal L}(m_{\nu}^{95})) =  ln({\cal L}^{MAX}) - 1.92 $
(this method is almost equivalent to the one based on the integration of 
the likelihood, used for example by CLEO, when the likelihood shape is 
fairly Gaussian as in this case).

To estimate the systematic error all the {\it modified} likelihoods containing
the effect of the different systematic sources would be needed. Since 
they are not published a rough estimate of the systematics have been obtained 
by multiplying each likelihood by a constant factor which brings, for each 
channel, the limit on $\mnu$ to be equal to the ones which includes the systematics. 
The total systematics has been obtained as followed: the systematic error for each 
combined mode is obtained by subtracting to the limit derived with 
the modified likelihoods the one obtained without systematics; all these 
errors are added in quadrature (in this way the possible correlations between 
the different systematic errors are not taken into account) giving a 
total systematics of 1.4 MeV.
By adding linearly this error the the statistical result a combined 
ALEPH+OPAL 95$\%$ C.L upper limit of 15 MeV on $\mnu$ is obtained. 
I want to stress that 
this combination is unofficial and approximated. The aim is to give an idea
of the gain which could be achieved with the combination of the LEP results 
and to push the ALEPH, DELPHI and OPAL collaboration to produce an official
combined $\mnu$ limit.

\section{Comparison with CLEO results}

\begin{figure}[h]      
\centerline{\epsfxsize 2.0 truein \epsfbox{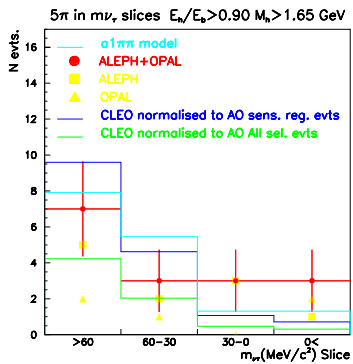}}   
\vskip 1.0 cm
\caption[]{
\label{cleovslep}
\small The LEP (ALEPH+OPAL) five-prong events compared with the CLEO ones in terms 
of $\mnu$ slices in the $\ehmh$ plane after requiring $M_{h}>1.65$ GeV and 
$E_{h}/E_{b}>0.9$. The red dots represents the LEP events. The dark-blue histogram 
shows the CLEO events normalised to the number of LEP events in this {\it sensitive} 
region. The green histogram shows the CLEO events normalised to the total number of 
LEP selected events. The ligh-blue histogram shows the expected events obtained by using 
a $\pi \pi a_{1}$ model for the five-prong tau decay.}
\end{figure}

The CLEO experiment has collected a huge statistics of tau decays 
at a centre-of-mass energy close to the 
$\Upsilon (4s)$ resonance and is expected to have a sensitivity to $\mnu$ larger than 
that of the LEP experiments. 
The performance of the CLEO~\cite{cleo5}, ALEPH~\cite{aleph5} and 
OPAL~\cite{opal5} 5$\pi$ analyses are compared in table~I. 
The CLEO limits are worse than the ALEPH one even though the CLEO statistics is a factor of 
five larger. This brings to the question: is CLEO unlucky or the are LEP results lucky ? 
It would be nice to evaluate for each experiment the {\it expected} limit on $\mnu$.
Its comparison with the actual one will tell us who is lucky and who is unlucky. 
Unfortunately the unknown hadronic dynamics doesn't allow the evaluation of the 
{\it a priori} sensitivity of an experiment. 
The ALEPH experiment claims that the probability to get such a lucky 
distribution in the $\ehmh$ plane is at the level of 15$\%$ if a 
model of the dynamics driven by $\pi \pi a_{1}$ is assumed in 5$\pi$ tau decays. 
At the same time CLEO claims that the probability to get a limit on $\mnu$ such a bad or worse 
than what they have obtained is at the level of 23$\%$ if a softer mass spectrum 
is assumed in the five-prong tau decays.
So the puzzle stays unsolved. 
What can be done is to compare data with data in the region where 
they are more sensitive to the tau-neutrino mass. This exercise is shown in 
Fig.~\ref{cleovslep} where the number of 5$\pi$ events are plotted in slices of 
iso-$\mnu$ in the $\ehmh$ plane;
in order to have more statistics the ALEPH and the OPAL events have been summed 
up in this comparison.

Only the events in a {\it sensitive} region which corresponds to $M_{h} >1.65$ GeV 
and $E_{h} / E_{b}>0.9$ are shown. The selection efficiency is assumed 
to be flat in the full $\ehmh$ plane. 
It finds out from this plot that the shapes of the LEP and CLEO data are compatible. 
This is shown by the comparison between the dark-blue line and the red dots 
of Fig.~\ref{cleovslep}; the CLEO events in the {\it sensitive} region have been 
normalised the number of ALEPH+OPAL events in the same region.

For what concerns the fraction of 5$\pi$ events selected in this region with respect 
to the total number of selected 5 $\pi$ events the compatibility is less good as can 
be observed by comparing the green line in  Fig.~\ref{cleovslep} with the red dots;
the ALEPH and the OPAL experiments select a total of 16 events in this {\it sensitive} 
region which should be compared with about 7 which is the number of events observed 
by CLEO in this {\it sensitive} region rescaled to the ALEPH+OPAL 5$\pi$ statistics. 
These two numbers are barely compatible. The number of 
{\it expected} events normalised to the ALEPH plus OPAL statistics is of about 18 
with the $\pi \pi a_1$ dynamics (this is showed by the light-blue line of Fig.~\ref{cleovslep}) 
and of about 9 with a softer dynamics similar to phase-space.

Even more intriguing is the fact that the likelihood shown by the CLEO
experiment at the TAU98 workshop~\cite{tau98} shows a peak at $\mnu \sim 18$ MeV. 
This likelihood is preliminary and doesn't include the systematic errors. 
If the ALEPH method described above 
is applied to this likelihood the value of $\mnu =0$ is excluded at more than 90 $\%$ 
confidence level. This could mean that CLEO is on the verge of a very interesting result 
or that (more probably but less interesting) there is in the CLEO likelihood a bias 
towards large neutrino masses. 
This bias would make the CLEO limit more conservative (explaining why their limit is so 
unlucky) and is therefore not warring in terms of the validity of their $\mnu$ 
upper limit. 
One should remember that most of the systematics determined by the 
different experiments are studied in terms of bias towards a massless tau-neutrino 
while less attention is played to possible sources which can mimic a massive neutrinos.
A typical example is the fact that all the experiments reduce the {\it dangerous} background 
(the one which can mimic a massless tau-neutrino) at the 1-2$\%$ level while backgrounds as 
high as $10 \%$ from higher decay multiplicities (like a $3 \pi^{\pm} \pi^{0}$ 
reconstructed as a $3 \pi^{\pm}$ tau decays) which can mimic a massive neutrino are accepted.
The CLEO experiment as still a large fraction of its statistics to analyse so I think 
that this intriguing situation will be clarified soon.

\begin{table}[h]
\caption{
Comparison between the performance of LEP and CLEO 5$\pi$ analyses.
For the different analyses the mass resolution in Mev, the energy 
resolution divided by the beam energy, the number of selected events, 
the efficiency in per cent and the 95 $\%$ 
upper limit, statistical only, on $\mnu$ in MeV are reported.}
\begin{tabular}{lccccc} 
 &  $\sigma (M_{h})$ GeV & $\sigma (E_{h})/E_{beam}$  & N evts & Efficiency $\%$ & $m_{\nu}^{95}$ MeV \\  
 \tableline 
ALEPH 5$\pi^{\pm}$            & 10-15 & 0.35/45.6 &  52 & 27   & 22.3 \\  
OPAL 5$\pi^{\pm}$             & 20-25 & 0.5/45.6  &  22 &  9   & 39.6 \\  
CLEO 5$\pi^{\pm}$             &  15   & 0.025/10. & 266 &  3   & 31.0 \\  
CLEO 3$\pi^{\pm}$2$\pi^{0}$   &  25   & 0.05/10.  & 207 &  0.4 & 33.0 \\
\end{tabular}
\end{table}

\section{Conclusions and Acknowledgements}
The LEP experiments have put constraints on
$\mnu$ by fitting the $\ehmh$ distribution in three- and five-prong tau decays.
The best results obtained by a single experiment is given by ALEPH 
which obtains $\mnu < 18.2$ MeV at 95$\%$ confidence level by combining 
the three- and five-prong results.  

An unofficial combination of the ALEPH and OPAL results shows that LEP
can exclude at 95$\%$ C.L. values of $\mnu$ higher than 15 MeV. 
In my personal opinion the CLEO experiment has the statistical 
power to go below this limit.

I want to thank Ronan McNulty from DELPHI, Achim Stahl from OPAL and 
Jean Duboscq from CLEO for the help that I received in preparing this talk.
A special thank goes to my ALEPH colleague (and friend) Luca Passalacqua
who shared with me three years of $\mnu$ measurements with the ALEPH
detector. I also want to thank the organisers of this very nice 
conference.

\end{document}